\documentclass[prl,amsfonts,twocolumn,showpacs,floatfix]{revtex4}
\usepackage{amsmath}
\usepackage{amsfonts}
\usepackage{amssymb}
\usepackage{bm}
\usepackage{txfonts}
\usepackage{tabularx}
\usepackage[dvips]{graphicx,color}

\newcommand{\TN}{T_{\mathrm{N}}}
\newcommand{\pcr}{PdCrO$_2$}
\newcommand{\pco}{PdCoO$_2$}

\newcommand{\RS}{R_\mathrm{S}}

\begin{document}
%%%%% Title %%%%%
\title{
Unconventional Anomalous Hall Effect in the Metallic Triangular-Lattice Magnet PdCrO$_{2}$
}
%%%%%%%%%%%%%%%%%

%%%%% Authors %%%%%
\author{Hiroshi Takatsu}
%\email{takatsu@scphys.kyoto-u.ac.jp}
\affiliation{Department of Physics, Graduate School of Science, Kyoto University, Kyoto 606-8502, Japan}

\author{Shingo Yonezawa}
\affiliation{Department of Physics, Graduate School of Science, Kyoto University, Kyoto 606-8502, Japan}

\author{Satoshi Fujimoto}
\affiliation{Department of Physics, Graduate School of Science, Kyoto University, Kyoto 606-8502, Japan}

\author{Yoshiteru Maeno}
\affiliation{Department of Physics, Graduate School of Science, Kyoto University, Kyoto 606-8502, Japan}
%%%%%%%%%%%%%%%%%%%
\date{\today}

%%%%% Abstract %%%%%
\begin{abstract}
We experimentally reveal an unconventional anomalous Hall effect (UAHE) 
in a quasi-two-dimensional triangular-lattice antiferromagnet PdCrO$_2$. 
%%%%%
Using high quality single crystals of PdCrO$_2$,
we found that the Hall resistivity $\rho_{xy}$ deviates
from the conventional behavior below $T^{*}\simeq20$~K, noticeably 
lower than $\TN=37.5$~K, at which Cr$^{3+}$ ($S = 3/2$) spins order 
in a 120$^\circ$ structure.
%%%%%
In view of the theoretical expectation that 
the spin chirality cancels out in the simplest 120$^\circ$ spin structure,
we discuss required conditions for the emergence of UAHE within Berry-phase
mechanisms.
\end{abstract}
%%%%%%%%%%%%%%%%%%%%

\pacs{72.15.-v, 75.47.-m, 75.47.Lx}
\maketitle

%%%%%%%%%%%%%%%%%%%%%%%%%%%%%%%%%%%%%%
%%%%%%%%%  Introduction %%%%%%%%%%%%%
%%%%%%%%%%%%%%%%%%%%%%%%%%%%%%%%%%%%%%
Recently, there has been a rapid progress in the study of 
unconventional magnetic phenomena which cannot be described 
solely in terms of the conventional order parameter, i.e.,
magnetization~\cite{S.W.CheongNature-Materials2007,Taguchi2001}.
%%%
One example of such phenomena is the unconventional anomalous Hall effect (UAHE)
in frustrated spin systems~\cite{Taguchi2001,Y.YasuiJPSJ2006,Machida2007,Y.MachidaNature2009,P.Matl1998PRB, J.YePRL1999,Ohgushi2000,TataraJPSJ2002,N.NagaosaJPSJ2006,D.Xiao2009,T.TomizawaPRB2009}, which cannot be accounted for by conventional 
AHE mechanisms based on spin-orbit interaction (SOI) to magnetization $M$~\cite{KarplusPR1954,J.SmitPhysica1955,L.BergerPRB1970}.
%%%
%Empirically,
The Hall resistivity $\rho_{xy}$ violates the empirical relation
expressed as a linear combination of terms proportional to
the magnetic induction $B=H+4\pi M$ and to $M$:
\begin{equation}
\rho_{xy}(H,T)=R_{0}(T)B +4\pi R_{\mathrm{S}}(T)M,
\label{eq.1}
\end{equation}
where $R_{0}$ and $R_{\mathrm{S}}$
are the ordinary Hall coefficient and the anomalous Hall coefficient,
respectively~\cite{C.M.Hurd}.
The first term originates from the Lorenz force and 
the second term is attributed to the orbital motion
of the spin polarized electrons by SOI.
%%%
For the origin of UAHE,
various new mechanisms including those based on multiple spin order 
parameters, namely spin chiralities, have been
proposed~\cite{J.YePRL1999,Ohgushi2000,TataraJPSJ2002,N.NagaosaJPSJ2006}.
%%%
These mechanisms are based on the Berry phase theory~\cite{Berry1984} 
taking into account a finite phase gain of the wavefunction of 
a conduction electron 
as it circulates through the field of 
the nontrivial (i.e. non-collinear and non-coplanar) spin structure.
%%%
The single-valuedness of the wavefunction enforces 
the conduction electrons to be subjected to a fictitious magnetic field, 
analogously to the Aharonov-Bohm effect~\cite{Aharonov1959}, 
leading to UAHE~\cite{J.YePRL1999,Ohgushi2000,TataraJPSJ2002,N.NagaosaJPSJ2006,D.Xiao2009,T.TomizawaPRB2009}.
%%%

%%%
Geometrically frustrated magnets are promising for 
investigating such UAHE,
because they often exhibit nontrivial 
spin configurations with non-vanishing spin chiralities. 
%%%
However,
the observation of UAHE has been limited to only a handful of bulk
materials with the three-dimensional (3D) analogue to the triangular
lattice (TL)~\cite{Taguchi2001,Machida2007}. 
%%%
Moreover, for the archetypal example of geometrically frustrated spin systems,
an antiferromagnet with a two-dimensional (2D) TL, UAHE has not been
experimentally reported nor theoretically expected~\cite{N.NagaosaJPSJ2006}.
%%%
It is thus important to find a conductive material with spins on a TL and 
investigate its Hall effect.
%%%
In this context,
we have studied the transport and magnetic properties of 
the metallic 2D-TL antiferromagnet PdCrO$_2$,
which should provide a unique testing ground 
for the clarification of the unresolved mechanism of UAHE.
%%%

PdCrO$_2$ crystallizes in the delafossite structure 
with the $R\bar{3}m$ symmetry
consisting of layers of Pd triangles and Cr triangles
stacking along the $c$ axis.
In this oxide,
localized spins of the Cr$^{3+}$ ions ($S=3/2$) exhibit
an antiferromagnetic order at $\TN = 37.5$~K, 
forming a 120$^{\circ}$ spin structure~\cite{Mekata1995,H.TakatsuPRB2009}.
This is one of the simplest spin structure among the structures
with the $\sqrt{3}\times\sqrt{3}$ periodicity consisting of 
three magnetic sublattices.
Metallic conductivity, maintained down to 
the lowest temperatures~\cite{H.TakatsuPRB2009},
is predominantly attributable to the Pd 4d$^9$ electrons, 
analogous to the isostructural non-magnetic metal PdCoO$_2$~\cite{Takatsu2007,V.EyertChmMater2008,H.J.NohPRL2009,K.Kim2009}.
%%%%%
Since majority of the known 2D-TL magnets are insulators or semiconductors,
PdCrO$_2$ is envisaged to serve as a standard for the Hall effect 
in metallic 2D-TL magnets with the 120$^\circ$ spin structure.
In this Letter,
we report a clear observation of unexpected UAHE in PdCrO$_2$
from the single-crystalline study.
We observed that $\rho_{xy}$ clearly deviates from the 
$H$--linear dependence and even changes its sign, although $M$
increase linearly with $H$.
This behavior sharply contrasts with the empirical behavior
expressed by Eq.~(\ref{eq.1}).
%%%%%%%%%

%%%%%%%%%%%%%%%%%%%%%%%%%%%%%%%%%%%%
%%%%%%%%%  Experiments %%%%%%%%%%%%%
%%%%%%%%%%%%%%%%%%%%%%%%%%%%%%%%%%%%
%%%%% Field dependence of the Hall resistivity and magnetization %%%%%
\begin{figure*}[t]
\begin{center}
 \includegraphics[width=0.7\textwidth]{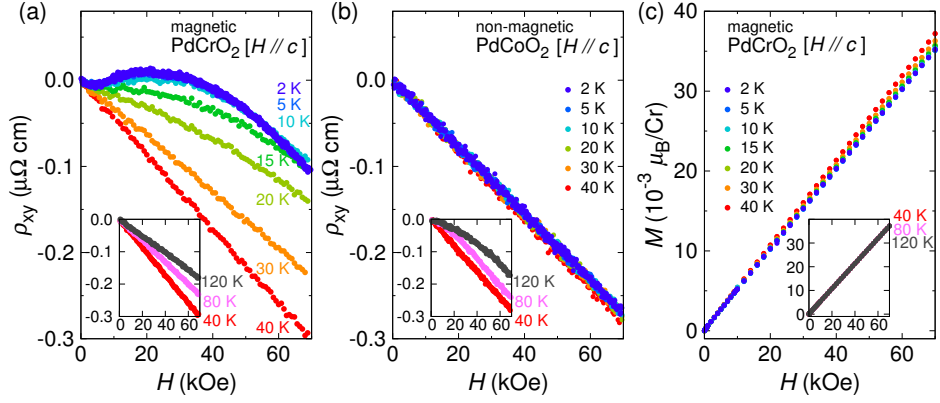}
\caption{
Field dependence of the Hall resistivity $\rho_{xy}$ for (a) PdCrO$_2$
($T_\mathrm{N}=37.5$~K) and for (b) non-magnetic PdCoO$_2$. 
(c) Field dependence of the magnetization $M$ 
for PdCrO$_2$.
Once the temperature decreases below $T^{*}\simeq20$~K,
$\rho_{xy}$ of PdCrO$_2$ strongly deviates from it and exhibits
a hump to positive values at around 10--30~kOe. 
}
\label{fig.1}
\end{center}
\end{figure*}
%%%%%%%%%%%%%%%%%%%%%%%%%%%%%%%%%%%%%%%%%%%%%%%%%%%%%%%%%%%%%%%%%%%%%%%%
Single crystals of PdCrO$_2$ were grown by a flux method
and characterized with the powder x-ray diffraction 
and energy dispersive x-ray analysis~\cite{Single_PdCrO2}.
%%%%%%%%%
The magnetoresistivity $\rho_{xx}$ and the Hall resistivity $\rho_{xy}$,
evaluated by reversing the field direction,
were simultaneously measured with a dc four-probe method 
with six contacts; the magnetic field $H$ was applied along the $c$ axis
($\langle001\rangle$ direction) and the current $I$ was applied 
in the $ab$ plane ($\langle110\rangle$ direction).
In order to extract effects of the frustrated spins,
the data for non-magnetic PdCoO$_2$ were compared.
%%%%%%%%%
The dc magnetization $M$ of PdCrO$_2$ in fields along the $c$ axis and 
in the $ab$ plane were measured with a SQUID magnetometer (Quantum Design MPMS)
for samples consisting of aligned crystals.

%%%%%%%%%%%%%%%%%%%%%%%%%%%%%%%%
%%%%%%%%%  Results %%%%%%%%%%%%%
%%%%%%%%%%%%%%%%%%%%%%%%%%%%%%%%
Figures \ref{fig.1} (a) and (b) represent the field dependence of $\rho_{xy}$ 
of \pcr\ and \pco\ measured at several temperatures.
For \pcr,\ 
$\rho_{xy}$ exhibits a linear field dependence above and near $\TN$
with a negative slope, indicating the dominance of electron-like carriers.
With decreasing temperature below $\TN$,
the slope rapidly changes with the magnetic phase transition.
%%%%%
Curiously, an unusual non-linear field dependence with 
a hump around 10--30~kOe emerges at temperatures below $T^{*}\simeq20$~K.
This behavior is reproducibly observed in different crystals we investigated.
%%%%%
In contrast,
$\rho_{xy}$ of non-magnetic PdCoO$_2$ exhibits a linear field
dependence without a slope change below 40~K.
%%%%%
We note here that 
the slight non-linearity of $\rho_{xy}$ of PdCoO$_2$ at elevated temperatures
(inset of Fig.~\ref{fig.1}(b)) is attributable to mechanisms such as multi-band effects 
or the scattering by optical phonons.
%%%%%
These results clearly indicate that the localized Cr spins affects
$\rho_{xy}$ of PdCrO$_2$.
%%%%%%%%%%%%%

Let us compare $\rho_{xy}$ and $M$ of PdCrO$_2$ (Fig.~\ref{fig.1}(c)).
%%%%%
Above $\TN$, 
both $\rho_{xy}$ and $M$ exhibit a linear field dependence,
and the relation ({\ref{eq.1}}) holds.
%%%%%
Between $\TN$ and $T^{*}$, 
the relation ({\ref{eq.1}}) still holds.
The slope change of $\rho_{xy}(H, T)$ 
in these temperature regions can be interpreted 
within the conventional AHE behavior, i.e.,
the temperature dependence of $R_{\mathrm{S}}$.
However, 
once the temperature decreases below $T^{*}$, 
$\rho_{xy}$ deviates from the linearity 
although $M$ still varies linearly on the magnetic fields. 
This result manifests that UAHE appears at 
temperatures below $T^{*}$.
%%%%%%%%%%%%%

\begin{figure}[t]
\begin{center}
 \includegraphics[width=0.41\textwidth]{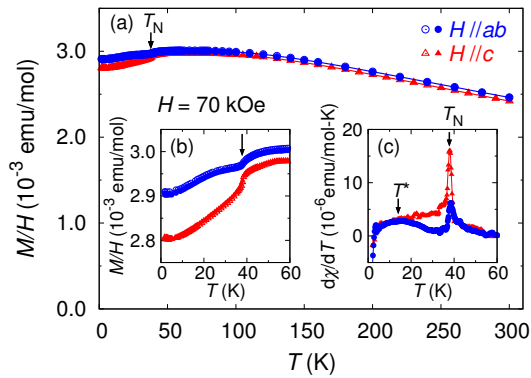}
\caption{
(a) Temperature dependence of the magnetic susceptibility 
of single crystalline PdCrO$_2$ 
for fields along the $c$ axis ($\chi_{c}=M_{c}/H$) and 
in the $ab$ plane ($\chi_{ab}=M_{ab}/H$) at 70~kOe. 
(b) Details of the anisotropy below $\TN$.
Open (closed) symbols represent data in
the field cooling (zero-field cooling) condition.
(c) Temperature derivative of $\chi_{c}$ and $\chi_{ab}$.
}
\label{fig.2}
\end{center}
\end{figure}

%%%%%%%%%%%%%%%%%%%%%%%%%%%%%%%%%%%%%%%%%%%%%%%%%%%%%%%%%%%%%%%%%%%%%%%
%\section{magnetic susceptibility of PdCrO$_2$}
%%%%%%%%%%%%%%%%%%%%%%%%%%%%%%%%%%%%%%%%%%%%%%%%%%%%%%%%%%%%%%%%%%%%%%%
Figure~\ref{fig.2}(a) compares the  temperature dependence of the magnetic
susceptibility $\chi\,(=M/H)$ with applied magnetic fields along 
the $c$ axis and in the $ab$ plane.
We confirmed that $\chi$ is isotropic above $\TN$
within the experimental precision of $5\times10^{-5}$~emu/mol between the measurements.
This result indicates that PdCrO$_2$ constitutes the Heisenberg spin system.
In contrast, $\chi$ becomes anisotropic below $\TN$ 
with a sharp drop in $\chi_c$ (Fig.~\ref{fig.2} (b)). 
Combined with the 120$^\circ$ spin structure 
determined from neutron diffraction,
such anisotropy in PdCrO$_2$ indicates that the spins order 
in a plane containing the $c$ axis,
and that they are coupled antiferromagnetically between 
the layers~\cite{T.HironeJPSJ1957}.
Interestingly, 
a broad maximum appears around $T^{*}$
in $\mathrm{d}\chi_{ab}/\mathrm{d}T$ (Fig.~\ref{fig.2}(c)).
This implies a minute modification of the magnetic structure.
It should be noted that the specific heat in our previous polycrystalline 
study~\cite{H.TakatsuPRB2009}, as well as in our recent single-crystalline study, 
also revealed a small anomaly around $T^{*}$.
%%%%%%%%%%%%%

%%%%%%%%%%%%%%%%%%%%%%%%%%%%%%%%%%%%%%%%%%%%%%%%%%%%%%%%%%%%%%%%%%%%%%%
%\section{Temp. dependence of $R_{s}$}
%%%%%%%%%%%%%%%%%%%%%%%%%%%%%%%%%%%%%%%%%%%%%%%%%%%%%%%%%%%%%%%%%%%%%%%
In order to express the unconventional nature of the AHE, 
let us extend Eq.~(\ref{eq.1}) to allow the anomalous Hall 
coefficient $R_{\mathrm{S}}$ to depend on field as well as on temperature:
\begin{alignat}{1}
R_{\mathrm{S}}(H, T)=\{\,\rho_{xy}(H,T)-R_0B \}/4\pi M(H,T).
\label{eq.2}
\end{alignat}
Here we estimate the ordinary Hall resistivity $R_0B$ of PdCrO$_2$
using $R_0$ of PdCoO$_2$, since PdCoO$_2$ is a good non-magnetic reference 
system to PdCrO$_2$ for the following reasons:
The values of the electronic specific heat coefficient~\cite{Takatsu2007,H.TakatsuPRB2009}, 
of room-temperature in-plane resistivity~\cite{Takatsu2007,H.TakatsuJPCS2010}, 
and of the Hall resistivity above $\TN$ are nearly the same for both compounds;
the band structure calculation revealed similar electronic configurations and Fermi 
surfaces~\cite{Shishidou}.
In the analysis, 
$R_0$ of PdCoO$_2$ is represented by the value at 2~K, $R_0=-3.86\times10^{-4}$~cm$^3$/C,
since it is indeed temperature independent below 40~K. 
%%%%%
Furthermore, $B$ in PdCrO$_2$ is approximated by the applied magnetic field,
$B=H+4\pi M\simeq H$, 
since $4\pi M$ is only about 0.1\% of $H$ 
in the measured field region.
%%%%%
Figure~4 shows the temperature dependence of $R_\mathrm{S}$.
It is clear that $R_\mathrm{S}$ indeed becomes field dependent only 
below $T^{*}$,
demonstrating the violation of Eq.~(\ref{eq.1}).

%%%%% Temperature dependence of the magnetic susceptibility %%%%%
%%%%%%%%%%%%%%%%%%%%%%%%%%%%%%%%%%%%%%%%%%%%%%%%%%%%%%%%%%%%%%%%%%%%%%%%

%%%%%%%%%%%%%%%%%%%%%%%%%%%%%%%%%%%%%%%%%%%%%%%%%%%%%%%%%%%%%%%%%%%%%%%
%\section{Finite $R_{s}$}
%%%%%%%%%%%%%%%%%%%%%%%%%%%%%%%%%%%%%%%%%%%%%%%%%%%%%%%%%%%%%%%%%%%%%%%
For conventional clean magnetic conductors, 
$\RS$ is known to diminish at low temperatures~\cite{C.M.Hurd}.
It is remarkable that $\RS$ of PdCrO$_2$ retains 
a large value at low temperatures, although it is a clean metal with
the estimated mean free path of 30~$\mu$m~\cite{Single_PdCrO2}. 
This provides additional evidence for UAHE of PdCrO$_2$. 
Such unconventional field and temperature dependence of $\RS$
is also observed in conductive pyrochlore magnets with nontrivial 
spin structures~\cite{Taguchi2001, Machida2007}.

%%%%%%%%%%%%%%%%%%%%%%%%%%%%%%%%%%%%%%%%%%%%%%%%%%%%%%%%%%%%%%%%%%%%%%%
%\section{Anomaly at $T^{*}$ in $\sigma_{xy}$}
%%%%%%%%%%%%%%%%%%%%%%%%%%%%%%%%%%%%%%%%%%%%%%%%%%%%%%%%%%%%%%%%%%%%%%%
For theoretical analyses,
a quantity of more fundamental physical significance
is the Hall conductivity $\sigma_{xy}$,
evaluated from $\rho_{xy}$ and $\rho_{xx}$ 
through the relation $\sigma_{xy}=\rho_{xy}/(\rho_{xx}^2+\rho_{xy}^2)$.
Figure~\ref{fig.4}(a) represents the overall field and 
temperature dependence of $\sigma_{xy}$. 
This 3D plot clearly indicates a trench representing $T^{*}$
and an island centered at about 15~kOe below 10~K.
This island represents the main characteristic of UAHE,
because, based on the chirality mechanism discussed below,
such an island corresponds to the opposite direction of 
the fictitious field with respect to the applied field direction.
%%%%%
The field dependence at low temperatures is more clearly shown in
Fig.~\ref{fig.4}(b).
The negative initial slope representing the Lorentz force term
is taken over by the unconventional contribution.
%%%%%

%%%%%%%%%%%%%%%%%%%%%%%%%%%%%%%%%%%
%%%%%%%%%  Conclusion %%%%%%%%%%%%%
%%%%%%%%%%%%%%%%%%%%%%%%%%%%%%%%%%%
We have revealed that the 2D-TL magnet PdCrO$_2$ exhibits UAHE, 
which cannot be ascribed solely by the conventional mechanism 
based on the spin-orbit coupling to the magnetization.
To the best of our knowledge,
this is the first UAHE report among the TL systems.
%%%%%
Moreover, the observed anomalies in $\mathrm{d}\chi_{ab}/\mathrm{d}T$
and the magnetic specific heat, as well as the comparison with
$\rho_{xy}$ of non-magnetic PdCoO$_2$, indicate that 
the unconventional behavior below $T^{*}$ 
is not related to Fermi surface effects
such as the magnetic breakdown~\cite{C.M.Hurd}.
%%%%%
Instead, the observed unconventional feature is similar to those
reported in other 3D geometrically frustrated spin 
systems~\cite{Taguchi2001, Machida2007}.
%%%%%
It is therefore promising to pursue the connection with 
various Berry-phase mechanisms in this 2D-TL system as well.
%%%%%

%%%%%
The scalar spin chirality mechanism
is one of the candidates to explain UAHE in frustrated magnets.
Let us consider
the local exchange field acting on Pd sites and the scalar spin chirality
$\chi^{\mathrm{Pd}}_{ijk}= \bm{s}_{i}\cdot(\bm{s}_{j}\times \bm{s}_{k})$.
Here, $\bm{s}_{i}$ is the conduction electron spin at the Pd site $i$; 
we assumed that $\bm{s}_{i}$ tends to align 
in the direction of the local exchange field 
$\bm{I}_{i}=\sum_{l=1}^{6}J\bm{S}_{l}$, created by Cr spins in the layers below
and above the Pd site.
$J$ is the coupling between the hopping Pd $d$-electrons and the 
localized Cr $d$-electrons, and $\bm{S}_{l}$ is a Cr$^{3+}$ spin 
at a site $l$ surrounding the Pd ion at the site $i$ (Fig.~\ref{fig.5}(a)).
Such a model has also been used for the pyrochlore 
Pr$_2$Ir$_2$O$_7$~\cite{Machida2007}.
In this model,
$\chi^{\mathrm{Pd}}_{ijk}$ is zero for 
the simple 120$^\circ$ spin structure of Cr spins
because $\bm{I}_{i}=0$ everywhere.
In order to give rise to finite $\chi^{\mathrm{Pd}}_{ijk}$,
it is necessary that a modification of the spin structure occurs at $T^{*}$.
Such a modification is in fact anticipated by the observed weak anomalies in
the susceptibility and specific heat.
%%%%%

%%%%% Temperature dependence of the anomalous Hall resistivity estimated by subtracting the Lorenz term %%%%%
\begin{figure}[t]
\begin{center}
\includegraphics[width=0.41\textwidth]{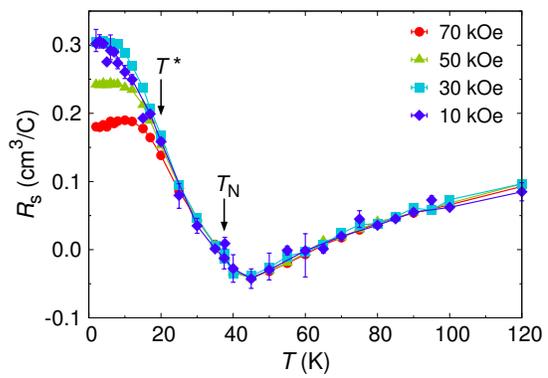}
\caption{
Temperature dependence of the anomalous Hall coefficient
$R_\mathrm{S}(H,T)$ represented by Eq.~(\ref{eq.2}).
$R_\mathrm{S}(H,T)$ does not depend on the applied magnetic field 
at temperatures above $T^{*}\simeq20$~K, whereas below $T^{*}$ 
it is field dependent, indicating the appearance of UAHE below $T^{*}$.
}
\label{fig.3}
\end{center}
\end{figure}
%%%%%%%%%%%%%%%%%%%%%%%%%%%%%%%%%%%%%%%%%%%%%%%%%%%%%%%%%%%%%%%%%%%%%%%

%%%%%
With the limited size of the single crystals currently available,
it is technically difficult to determine the subtle spin modification 
below $T^{*}$. 
Nevertheless, we can provide a few prerequisites for 
the actual spin modification.
%%%%%
Under a magnetic field,
$\bm{I}_{i}$ becomes non-zero because 
of the polarization of the Cr spins surrounding a Pd ion.
%%%%%
As a first prerequisite, 
for $\chi^{\mathrm{Pd}}_{ijk}$ not to vanish, 
$\bm{I}_{m}$ ($m=i,j,k$) should be non-coplanar.
This requires that the Cr-spin configuration should be non-coplanar and
moreover break the $\sqrt{3}\times\sqrt{3}$ periodicity.
%%%%%
Secondly, $\chi^{\mathrm{Pd}}_{ijk}$ after averaged over the
entire lattice should not vanish.
By analogy with the Berry-phase theory of magnetic
nanostructures~\cite{BrunoPRL2004},
one can deduce that both $\Delta\theta_{ij}=\theta_{i}-\theta_{j}$ and
$\Delta\phi_{ij}=\phi_{i}-\phi_{j}$ should not vanish for the appearance of 
a net non-vanishing $\chi^{\mathrm{Pd}}_{ijk}$.
Here, $(\theta_{i}, \phi_{i})$ specifies the polar-coordinate 
direction of $\bm{I}_{i}$. 
These conditions require a somewhat complicated 
modulation of the Cr-spin configuration. 

%%%%%
Based on these prerequisites,
let us examine possible modifications of the spin structure
in PdCrO$_2$. 
%%%%%
We should first note that
as long as the $\sqrt{3}\times\sqrt{3}$ periodicity is maintained,
simple modifications such as a change from
antiferromagnetic to ferromagnetic interlayer coupling and 
a spin flop to the so-called ``up-up-down'' structure~\cite{KawamuraJPSJ1985} 
do not satisfy the first prerequisite.
Such a spin flop is indeed inconsistent with the absence of 
a magnetization plateau in Fig.~\ref{fig.1}(c).
%%%%%
A simple illustrative example satisfying both prerequisites
is a spin structure in which the normal vector of the 120$^\circ$ spin plane
precesses with the secondary propagation vector $\bm{q}_1=(1/3, 2/3, 0)$.
In zero field, and in the case of an antiferromagnetic inter-Cr-layer stacking 
indicated by the susceptibility data, the total chirality averages to zero in the magnetic unit cell.
However, additional Cr spin polarization under a $c$-axis magnetic field leads to a non-vanishing net 
chirality through a breaking of symmetry between the two Cr layers, as shown in Fig.~\ref{fig.5}(b).
%%%%%
We note that the orbital Berry phase mechanism~\cite{T.TomizawaPRB2009}
may be additionally relevant because of
the multi-band conductivity of PdCrO$_2$~\cite{Shishidou}.
%%%%%

%%%%% Temperature dependence of the Hall conductivity %%%%%
\begin{figure}[t]
\begin{center}
 \includegraphics[width=0.33\textwidth]{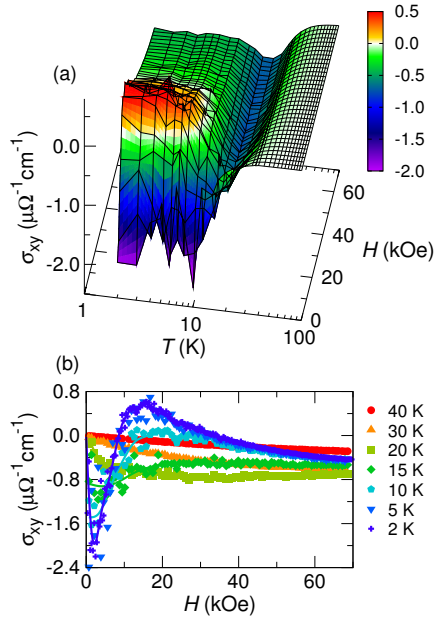}
\caption{
(a) 3D representation of the overall field and temperature 
dependence of the the Hall conductivity $\sigma_{xy}$.
(b) Field dependence of $\sigma_{xy}$ at temperatures below 40~K.
From a point of view based on the scalar spin chirality mechanism,
the fictitious field has the opposite sign to the applied magnetic field
in the red region in (a).
}
\label{fig.4}
\end{center}
\end{figure}
%%%%%%%%%%%%%%%%%%

%%%%% Temperature dependence of the Hall conductivity %%%%%
\begin{figure}[t]
\begin{center}
 \includegraphics[width=0.38\textwidth]{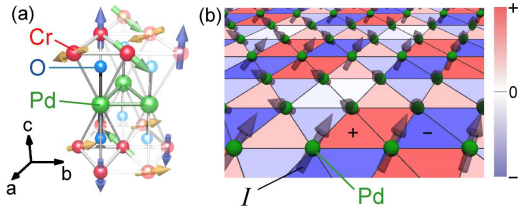}
\caption{
(a) Positions of Pd and Cr ions in PdCrO$_2$.
The arrows represent Cr spins in a 120$^\circ$ spin structure
with the antiferromagnetic interlayer coupling.
(b) Calculated spin chirality $\chi^{\mathrm{Pd}}_{ijk}$ mapping 
in the Pd-TL net for the modulated Cr spin structure 
in which the normal vector of the 120$^\circ$ plane precesses. 
The arrows indicate the exchange field $\bm{I}$ on Pd ions;
the red and blue colorings of triangles represent plus and minus 
values of $\chi^{\mathrm{Pd}}_{ijk}$, respectively.
Slight tilt of spins due to the magnetic field leads to overall 
non-zero chirality.
The parameters used in this figure are 
the precession angle of 5$^\circ$ and
the magnetic field strength of 70~kOe.
The chirality values are nonlinear in these parameters.
}
\label{fig.5}
\end{center}
\end{figure}
%%%%%%%%%%%%%%%%%%
%%%%
In summary, the anomalous Hall effect in a 2D-TL antiferromagnet PdCrO$_2$
becomes unconventional below 20~K, substantially lower than $\TN$.
%%%%%
It is remarkable that 
such UAHE indeed emerges in a simplest geometrically frustrated system,
namely the Heisenberg spins on a 2D-TL interacting with  conduction electrons.
For this reason, it is expected that
PdCrO$_2$ serves as an archetypal system toward clarification of 
the unresolved mechanism of UAHE.
%%%%% 
Detailed studies on the magnetic and crystal structure below $T^{*}$
are important in the future for deeper understandings of the origin of UAHE.

%%%%% 

%%%%%%%%%%%%%%%%%%%%%%%%%%%
%%%%% Acknowledgement %%%%%
%%%%%%%%%%%%%%%%%%%%%%%%%%%
We acknowledge K. Ishida, M. Kriener, C. Michioka, Y. Nakai, S. Kittaka,
J. Sobota and Z. X. Shen for their supports and
H. Kontani, T. Tomizawa, G. Tatara, T. Shishidou and T. Oguchi 
for fruitful discussion.
This work was supported by the Grant-in-Aid for the Global COE Program 
``The Next Generation of Physics, Spun from Universality and Emergence'' 
from the Ministry of Education, Culture, Sports, Science and Technology (MEXT) of 
Japan and a Grant-in-Aid for Scientific Research from the Japan Society for the Promotion of Science (JSPS).
H.T. is financially supported as a JSPS research fellow.

\bibliography{reference}

\begin{thebibliography}{30}
\expandafter\ifx\csname natexlab\endcsname\relax\def\natexlab#1{#1}\fi
\expandafter\ifx\csname bibnamefont\endcsname\relax
  \def\bibnamefont#1{#1}\fi
\expandafter\ifx\csname bibfnamefont\endcsname\relax
  \def\bibfnamefont#1{#1}\fi
\expandafter\ifx\csname citenamefont\endcsname\relax
  \def\citenamefont#1{#1}\fi
\expandafter\ifx\csname url\endcsname\relax
  \def\url#1{\texttt{#1}}\fi
\expandafter\ifx\csname urlprefix\endcsname\relax\def\urlprefix{URL }\fi
\providecommand{\bibinfo}[2]{#2}
\providecommand{\eprint}[2][]{\url{#2}}

\bibitem[{\citenamefont{Cheong and
  Mostovoy}(2007)}]{S.W.CheongNature-Materials2007}
\bibinfo{author}{\bibfnamefont{S.-W.} \bibnamefont{Cheong}} \bibnamefont{and}
  \bibinfo{author}{\bibfnamefont{M.}~\bibnamefont{Mostovoy}},
  \bibinfo{journal}{Nat. Mater.} \textbf{\bibinfo{volume}{6}},
  \bibinfo{pages}{13} (\bibinfo{year}{2007}).

\bibitem[{\citenamefont{Taguchi et~al.}(2001)\citenamefont{Taguchi, Oohara,
  Yoshizawa, Nagaosa, and Tokura}}]{Taguchi2001}
\bibinfo{author}{\bibfnamefont{Y.}~\bibnamefont{Taguchi}},
  \bibinfo{author}{\bibfnamefont{Y.}~\bibnamefont{Oohara}},
  \bibinfo{author}{\bibfnamefont{H.}~\bibnamefont{Yoshizawa}},
  \bibinfo{author}{\bibfnamefont{N.}~\bibnamefont{Nagaosa}}, \bibnamefont{and}
  \bibinfo{author}{\bibfnamefont{Y.}~\bibnamefont{Tokura}},
  \bibinfo{journal}{Science} \textbf{\bibinfo{volume}{291}},
  \bibinfo{pages}{2573} (\bibinfo{year}{2001}).

\bibitem[{\citenamefont{Yasui et~al.}(2006)\citenamefont{Yasui, Kageyama,
  Moyoshi, Soda, Sato, and Kakurai}}]{Y.YasuiJPSJ2006}
\bibinfo{author}{\bibfnamefont{Y.}~\bibnamefont{Yasui}},
  \bibinfo{author}{\bibfnamefont{T.}~\bibnamefont{Kageyama}},
  \bibinfo{author}{\bibfnamefont{T.}~\bibnamefont{Moyoshi}},
  \bibinfo{author}{\bibfnamefont{M.}~\bibnamefont{Soda}},
  \bibinfo{author}{\bibfnamefont{M.}~\bibnamefont{Sato}}, \bibnamefont{and}
  \bibinfo{author}{\bibfnamefont{K.}~\bibnamefont{Kakurai}},
  \bibinfo{journal}{J. Phys. Soc. Jpn.} \textbf{\bibinfo{volume}{75}},
  \bibinfo{pages}{084711} (\bibinfo{year}{2006}).

\bibitem[{\citenamefont{Machida et~al.}(2007)\citenamefont{Machida, Nakatsuji,
  Maeno, Tayama, Sakakibara, and Onoda}}]{Machida2007}
\bibinfo{author}{\bibfnamefont{Y.}~\bibnamefont{Machida}},
  \bibinfo{author}{\bibfnamefont{S.}~\bibnamefont{Nakatsuji}},
  \bibinfo{author}{\bibfnamefont{Y.}~\bibnamefont{Maeno}},
  \bibinfo{author}{\bibfnamefont{T.}~\bibnamefont{Tayama}},
  \bibinfo{author}{\bibfnamefont{T.}~\bibnamefont{Sakakibara}},
  \bibnamefont{and} \bibinfo{author}{\bibfnamefont{S.}~\bibnamefont{Onoda}},
  \bibinfo{journal}{Phys. Rev. Lett.} \textbf{\bibinfo{volume}{98}},
  \bibinfo{pages}{057203} (\bibinfo{year}{2007}).

\bibitem[{\citenamefont{Machida et~al.}(2009)\citenamefont{Machida, Nakatsuji,
  Onoda, Tayama, and Sakakibara}}]{Y.MachidaNature2009}
\bibinfo{author}{\bibfnamefont{Y.}~\bibnamefont{Machida}},
  \bibinfo{author}{\bibfnamefont{S.}~\bibnamefont{Nakatsuji}},
  \bibinfo{author}{\bibfnamefont{S.}~\bibnamefont{Onoda}},
  \bibinfo{author}{\bibfnamefont{T.}~\bibnamefont{Tayama}}, \bibnamefont{and}
  \bibinfo{author}{\bibfnamefont{T.}~\bibnamefont{Sakakibara}},
  \bibinfo{journal}{Nature} \textbf{\bibinfo{volume}{463}},
  \bibinfo{pages}{210} (\bibinfo{year}{2009}).

\bibitem[{\citenamefont{Matl et~al.}(1998)\citenamefont{Matl, Ong, Yan, Li,
  Studebaker, Baum, and Doubinina}}]{P.Matl1998PRB}
\bibinfo{author}{\bibfnamefont{P.}~\bibnamefont{Matl}},
  \bibinfo{author}{\bibfnamefont{N.~P.} \bibnamefont{Ong}},
  \bibinfo{author}{\bibfnamefont{Y.~F.} \bibnamefont{Yan}},
  \bibinfo{author}{\bibfnamefont{Y.~Q.} \bibnamefont{Li}},
  \bibinfo{author}{\bibfnamefont{D.}~\bibnamefont{Studebaker}},
  \bibinfo{author}{\bibfnamefont{T.}~\bibnamefont{Baum}}, \bibnamefont{and}
  \bibinfo{author}{\bibfnamefont{G.}~\bibnamefont{Doubinina}},
  \bibinfo{journal}{Phys. Rev. B} \textbf{\bibinfo{volume}{57}},
  \bibinfo{pages}{10248} (\bibinfo{year}{1998}).

\bibitem[{\citenamefont{Ye et~al.}(1999)\citenamefont{Ye, Kim, Millis,
  Shraiman, Majumdar, and Tasanovic}}]{J.YePRL1999}
\bibinfo{author}{\bibfnamefont{J.}~\bibnamefont{Ye}},
  \bibinfo{author}{\bibfnamefont{Y.~B.} \bibnamefont{Kim}},
  \bibinfo{author}{\bibfnamefont{A.~J.} \bibnamefont{Millis}},
  \bibinfo{author}{\bibfnamefont{B.~I.} \bibnamefont{Shraiman}},
  \bibinfo{author}{\bibfnamefont{P.}~\bibnamefont{Majumdar}}, \bibnamefont{and}
  \bibinfo{author}{\bibfnamefont{Z.}~\bibnamefont{Tasanovic}},
  \bibinfo{journal}{Phys. Rev. Lett.} \textbf{\bibinfo{volume}{83}},
  \bibinfo{pages}{3737} (\bibinfo{year}{1999}).

\bibitem[{\citenamefont{Ohgushi et~al.}(2000)\citenamefont{Ohgushi, Murakami,
  and Nagaosa}}]{Ohgushi2000}
\bibinfo{author}{\bibfnamefont{K.}~\bibnamefont{Ohgushi}},
  \bibinfo{author}{\bibfnamefont{S.}~\bibnamefont{Murakami}}, \bibnamefont{and}
  \bibinfo{author}{\bibfnamefont{N.}~\bibnamefont{Nagaosa}},
  \bibinfo{journal}{Phys. Rev. B} \textbf{\bibinfo{volume}{62}},
  \bibinfo{pages}{6065} (\bibinfo{year}{2000}).

\bibitem[{\citenamefont{Tatara and Kawamura}(2002)}]{TataraJPSJ2002}
\bibinfo{author}{\bibfnamefont{G.}~\bibnamefont{Tatara}} \bibnamefont{and}
  \bibinfo{author}{\bibfnamefont{H.}~\bibnamefont{Kawamura}},
  \bibinfo{journal}{J. Phys. Soc. Jpn} \textbf{\bibinfo{volume}{71}},
  \bibinfo{pages}{2613} (\bibinfo{year}{2002}).

\bibitem[{\citenamefont{Nagaosa}(2006)}]{N.NagaosaJPSJ2006}
\bibinfo{author}{\bibfnamefont{N.}~\bibnamefont{Nagaosa}}, \bibinfo{journal}{J.
  Phys. Soc. Jpn.} \textbf{\bibinfo{volume}{75}}, \bibinfo{pages}{042001}
  (\bibinfo{year}{2006}).

\bibitem[{\citenamefont{Xiao et~al.}()\citenamefont{Xiao, Chang, and
  Niu}}]{D.Xiao2009}
\bibinfo{author}{\bibfnamefont{D.}~\bibnamefont{Xiao}},
  \bibinfo{author}{\bibfnamefont{M.-C.} \bibnamefont{Chang}}, \bibnamefont{and}
  \bibinfo{author}{\bibfnamefont{Q.}~\bibnamefont{Niu}},
  \bibinfo{note}{arXiv:0907.2021v1}.

\bibitem[{\citenamefont{Tomizawa and Kontani}(2009)}]{T.TomizawaPRB2009}
\bibinfo{author}{\bibfnamefont{T.}~\bibnamefont{Tomizawa}} \bibnamefont{and}
  \bibinfo{author}{\bibfnamefont{H.}~\bibnamefont{Kontani}},
  \bibinfo{journal}{Phys. Rev. B} \textbf{\bibinfo{volume}{80}},
  \bibinfo{pages}{100401} (\bibinfo{year}{2009}).

\bibitem[{\citenamefont{Karplus and Luttinger}(1954)}]{KarplusPR1954}
\bibinfo{author}{\bibfnamefont{R.}~\bibnamefont{Karplus}} \bibnamefont{and}
  \bibinfo{author}{\bibfnamefont{J.~M.} \bibnamefont{Luttinger}},
  \bibinfo{journal}{Phys. Rev.} \textbf{\bibinfo{volume}{95}},
  \bibinfo{pages}{1154} (\bibinfo{year}{1954}).

\bibitem[{\citenamefont{Smit}(1955)}]{J.SmitPhysica1955}
\bibinfo{author}{\bibfnamefont{J.}~\bibnamefont{Smit}},
  \bibinfo{journal}{Physica} \textbf{\bibinfo{volume}{21}},
  \bibinfo{pages}{877} (\bibinfo{year}{1955}).

\bibitem[{\citenamefont{Berger}(1970)}]{L.BergerPRB1970}
\bibinfo{author}{\bibfnamefont{L.}~\bibnamefont{Berger}},
  \bibinfo{journal}{Phys. Rev. B} \textbf{\bibinfo{volume}{2}},
  \bibinfo{pages}{4559} (\bibinfo{year}{1970}).

\bibitem[{\citenamefont{Hurd}(1972)}]{C.M.Hurd}
\bibinfo{author}{\bibfnamefont{C.~M.} \bibnamefont{Hurd}},
  \emph{\bibinfo{title}{The Hall effect in metals and alloys}}
  (\bibinfo{publisher}{Plenum Press}, \bibinfo{address}{New York},
  \bibinfo{year}{1972}).

\bibitem[{\citenamefont{Berry}(1984)}]{Berry1984}
\bibinfo{author}{\bibfnamefont{M.~V.} \bibnamefont{Berry}},
  \bibinfo{journal}{Proc. R. Soc. London Ser. A}
  \textbf{\bibinfo{volume}{392}}, \bibinfo{pages}{45} (\bibinfo{year}{1984}).

\bibitem[{\citenamefont{Aharonov and Bohm}(1959)}]{Aharonov1959}
\bibinfo{author}{\bibfnamefont{Y.}~\bibnamefont{Aharonov}} \bibnamefont{and}
  \bibinfo{author}{\bibfnamefont{D.}~\bibnamefont{Bohm}},
  \bibinfo{journal}{Phys. Rev.} \textbf{\bibinfo{volume}{115}},
  \bibinfo{pages}{485} (\bibinfo{year}{1959}).

\bibitem[{\citenamefont{Mekata et~al.}(1995)\citenamefont{Mekata, Sugino,
  Oohara, Oohara, and Yoshizawa}}]{Mekata1995}
\bibinfo{author}{\bibfnamefont{M.}~\bibnamefont{Mekata}},
  \bibinfo{author}{\bibfnamefont{T.}~\bibnamefont{Sugino}},
  \bibinfo{author}{\bibfnamefont{A.}~\bibnamefont{Oohara}},
  \bibinfo{author}{\bibfnamefont{Y.}~\bibnamefont{Oohara}}, \bibnamefont{and}
  \bibinfo{author}{\bibfnamefont{H.}~\bibnamefont{Yoshizawa}},
  \bibinfo{journal}{Physica B} \textbf{\bibinfo{volume}{213}},
  \bibinfo{pages}{221} (\bibinfo{year}{1995}).

\bibitem[{\citenamefont{Takatsu et~al.}(2009)\citenamefont{Takatsu, Yoshizawa,
  Yonezawa, and Maeno}}]{H.TakatsuPRB2009}
\bibinfo{author}{\bibfnamefont{H.}~\bibnamefont{Takatsu}},
  \bibinfo{author}{\bibfnamefont{H.}~\bibnamefont{Yoshizawa}},
  \bibinfo{author}{\bibfnamefont{S.}~\bibnamefont{Yonezawa}}, \bibnamefont{and}
  \bibinfo{author}{\bibfnamefont{Y.}~\bibnamefont{Maeno}},
  \bibinfo{journal}{Phys. Rev. B} \textbf{\bibinfo{volume}{79}},
  \bibinfo{pages}{104424} (\bibinfo{year}{2009}).

\bibitem[{\citenamefont{Takatsu et~al.}(2007)\citenamefont{Takatsu, Yonezawa,
  Mouri, Nakatsuji, Tanaka, and Maeno}}]{Takatsu2007}
\bibinfo{author}{\bibfnamefont{H.}~\bibnamefont{Takatsu}},
  \bibinfo{author}{\bibfnamefont{S.}~\bibnamefont{Yonezawa}},
  \bibinfo{author}{\bibfnamefont{S.}~\bibnamefont{Mouri}},
  \bibinfo{author}{\bibfnamefont{S.}~\bibnamefont{Nakatsuji}},
  \bibinfo{author}{\bibfnamefont{K.}~\bibnamefont{Tanaka}}, \bibnamefont{and}
  \bibinfo{author}{\bibfnamefont{Y.}~\bibnamefont{Maeno}}, \bibinfo{journal}{J.
  Phys. Soc. Jpn.} \textbf{\bibinfo{volume}{76}}, \bibinfo{pages}{104701}
  (\bibinfo{year}{2007}).

\bibitem[{\citenamefont{Eyert et~al.}(2008)\citenamefont{Eyert, Fr{\'{e}}sard,
  and Maignan}}]{V.EyertChmMater2008}
\bibinfo{author}{\bibfnamefont{V.}~\bibnamefont{Eyert}},
  \bibinfo{author}{\bibfnamefont{R.}~\bibnamefont{Fr{\'{e}}sard}},
  \bibnamefont{and} \bibinfo{author}{\bibfnamefont{A.}~\bibnamefont{Maignan}},
  \bibinfo{journal}{Chem. Mater.} \textbf{\bibinfo{volume}{20}},
  \bibinfo{pages}{2370} (\bibinfo{year}{2008}).

\bibitem[{\citenamefont{Noh et~al.}(2009)\citenamefont{Noh, Jeong, Cho, Kim,
  Kim, Min, and Kim}}]{H.J.NohPRL2009}
\bibinfo{author}{\bibfnamefont{H.-J.} \bibnamefont{Noh}},
  \bibinfo{author}{\bibfnamefont{J.}~\bibnamefont{Jeong}},
  \bibinfo{author}{\bibfnamefont{J.~J. E.-J.} \bibnamefont{Cho}},
  \bibinfo{author}{\bibfnamefont{S.~B.} \bibnamefont{Kim}},
  \bibinfo{author}{\bibfnamefont{K.}~\bibnamefont{Kim}},
  \bibinfo{author}{\bibfnamefont{B.~I.} \bibnamefont{Min}}, \bibnamefont{and}
  \bibinfo{author}{\bibfnamefont{H.~D.} \bibnamefont{Kim}},
  \bibinfo{journal}{Phys. Rev. Lett.} \textbf{\bibinfo{volume}{102}},
  \bibinfo{pages}{256404} (\bibinfo{year}{2009}).

\bibitem[{\citenamefont{Kim et~al.}(2009)\citenamefont{Kim, Chul, and
  Min}}]{K.Kim2009}
\bibinfo{author}{\bibfnamefont{K.}~\bibnamefont{Kim}},
  \bibinfo{author}{\bibfnamefont{H.}~\bibnamefont{Chul}}, \bibnamefont{and}
  \bibinfo{author}{\bibfnamefont{B.~I.} \bibnamefont{Min}},
  \bibinfo{journal}{Phys. Rev. B} \textbf{\bibinfo{volume}{80}},
  \bibinfo{pages}{035116} (\bibinfo{year}{2009}).

\bibitem[{\citenamefont{Takatsu and Maeno}()}]{Single_PdCrO2}
\bibinfo{author}{\bibfnamefont{H.}~\bibnamefont{Takatsu}} \bibnamefont{and}
  \bibinfo{author}{\bibfnamefont{Y.}~\bibnamefont{Maeno}},
  \bibinfo{note}{(unpublished)}.

\bibitem[{\citenamefont{Hirone and Adachi}(1957)}]{T.HironeJPSJ1957}
\bibinfo{author}{\bibfnamefont{T.}~\bibnamefont{Hirone}} \bibnamefont{and}
  \bibinfo{author}{\bibfnamefont{K.}~\bibnamefont{Adachi}},
  \bibinfo{journal}{J. Phys. Soc. Jpn.} \textbf{\bibinfo{volume}{12}},
  \bibinfo{pages}{156} (\bibinfo{year}{1957}).

\bibitem[{\citenamefont{Takatsu et~al.}(2010)\citenamefont{Takatsu, Yonezawa,
  Michioka, Yoshimura, and Maeno}}]{H.TakatsuJPCS2010}
\bibinfo{author}{\bibfnamefont{H.}~\bibnamefont{Takatsu}},
  \bibinfo{author}{\bibfnamefont{S.}~\bibnamefont{Yonezawa}},
  \bibinfo{author}{\bibfnamefont{C.}~\bibnamefont{Michioka}},
  \bibinfo{author}{\bibfnamefont{K.}~\bibnamefont{Yoshimura}},
  \bibnamefont{and} \bibinfo{author}{\bibfnamefont{Y.}~\bibnamefont{Maeno}},
  \bibinfo{journal}{J. Phys. Conf. Ser.} \textbf{\bibinfo{volume}{200}},
  \bibinfo{pages}{012198} (\bibinfo{year}{2010}).

\bibitem[{Shi()}]{Shishidou}
\bibinfo{note}{T. Shishidou and T. Oguchi, presented at the 65th annual meeting
  of the Physical Society of Japan (2010) 23aGJ-14.}

\bibitem[{\citenamefont{Bruno et~al.}(2004)\citenamefont{Bruno, Dugaev, and
  Taillefumier}}]{BrunoPRL2004}
\bibinfo{author}{\bibfnamefont{P.}~\bibnamefont{Bruno}},
  \bibinfo{author}{\bibfnamefont{V.~K.} \bibnamefont{Dugaev}},
  \bibnamefont{and}
  \bibinfo{author}{\bibfnamefont{M.}~\bibnamefont{Taillefumier}},
  \bibinfo{journal}{Phys. Rev. Lett.} \textbf{\bibinfo{volume}{93}},
  \bibinfo{pages}{096806} (\bibinfo{year}{2004}).

\bibitem[{\citenamefont{Kawamura and Miyashita}(1985)}]{KawamuraJPSJ1985}
\bibinfo{author}{\bibfnamefont{H.}~\bibnamefont{Kawamura}} \bibnamefont{and}
  \bibinfo{author}{\bibfnamefont{S.}~\bibnamefont{Miyashita}},
  \bibinfo{journal}{J. Phys. Soc. Jpn} \textbf{\bibinfo{volume}{54}},
  \bibinfo{pages}{4530} (\bibinfo{year}{1985}).

\end{thebibliography}
\end{document}